\documentclass[prd,amsfonts,twocolumn,nofootinbib,showpacs]{revtex4}

\usepackage{graphicx}

\newcommand{\be}{\begin{equation}}
\newcommand{\ee}{\end{equation}}
\newcommand{\bea}{\begin{eqnarray}}
\newcommand{\eea}{\end{eqnarray}}

\newcommand{\gapp}{\mathrel{\raise.3ex\hbox{$>$}\mkern-14mu
              \lower0.6ex\hbox{$\sim$}}}
\newcommand{\lapp}{\mathrel{\raise.3ex\hbox{$<$}\mkern-14mu
              \lower0.6ex\hbox{$\sim$}}}

\newcommand\lsim{\lesssim}
\newcommand\gsim{\gtrsim}

\renewcommand\({\left(}
\renewcommand\){\right)}
\renewcommand\[{\left[}
\renewcommand\]{\right]}

\newcommand\eq[1]{Eq.~(\ref{#1})}
\newcommand\eqs[2]{Eqs.~(\ref{#1}) and (\ref{#2})}
\newcommand\eqss[3]{Eqs.~(\ref{#1}), (\ref{#2}), and (\ref{#3})}

\newcommand\eqst[2]{Eqs.~(\ref{#1})--(\ref{#2})}

\newcommand\eqreff[1]{(\ref{#1})}

\newcommand\eqssref[3]{(\ref{#1}), (\ref{#2}) and (\ref{#3})}

\newcommand\pa{\partial}

\newcommand\mpl{M_{\rm P}}

\def\calp{{\cal P}}

\def\calpz{{\calp_\zeta}}

\newcommand\bfx{{\mathbf x}}

\newcommand\GeV{\,\mbox{GeV}}

\newcommand\Mpc{\,\mbox{Mpc}}

\newcommand\sub[1]{_{\rm #1}}

\newcommand\mone{^{-1}}
\newcommand\mtwo{^{-2}}
\newcommand\mthree{^{-3}}
\newcommand\mfour{^{-4}}

\newcommand\half{^{1/2}}

\newcommand\eph{\epsilon_H}

\pacs{26.35.+c, 98.80.Cq, 98.80.Ft \hfill CERN-PH-TH/2007-242}

\begin{document}

\title{Black hole formation  and slow-roll inflation}

\author{Kazunori Kohri$^{1}$, David H. Lyth$^{1}$ 
and Alessandro Melchiorri$^{2,3}$}


\affiliation{
$^1$ Physics Department, Lancaster University LA1 4YB, UK, \\
$^2$ Physics Department and Sezione INFN, University of Rome 
``La Sapienza'', Ple Aldo Moro 2, 00185, Rome, Italy\\
$^3$ CERN, Theory Division, Geneva 23,CH-1211, Switzerland}

\begin{abstract}
Black hole formation may occur if
the spectrum of the curvature perturbation $\zeta$ increases strongly as the
scale decreases.  As no such increase is observed on cosmological scales,
black hole formation requires  strongly positive running $n'$ of the spectral
index $n$, though the running might only kick in  below the 
`cosmological  scales' probed by the CMB anisotropy and galaxy surveys.
A concrete and well--motivated
 way of producing this running is through the running
mass model of slow roll inflation.
We obtain a new observational  bound $n'<0.026$ on the running provided
by this   model, improving an earlier result by a factor two.
We also discuss black hole production in more general scenarios.
We show that the usual conditions $\epsilon \ll 1$ and $| \eta | \ll 1$
are enough to derive the spectrum ${\cal P}_{\zeta}(k)$, the
introduction of higher order parameters $\xi^{2}$ etc. being  
optional.
\end{abstract}

\maketitle

\section{Introduction}

The primordial curvature perturbation $\zeta$
is only of order $10\mfour$ on 
cosmological scales, but it might be of order 1 on smaller scales.
 Primordial black holes will 
 then form as 
those scales enter the horizon, with possibly observable consequences.
The purpose of this paper is to see to what extent the value of order 1
is reasonable, taking into account observational constraints and current 
thinking about the origin of $\zeta$.

In Section II we see what is required for black hole formation, in terms of
the spectral index $n(\ln k)\equiv d\ln\calpz/d\ln k$, which specifies  
 the scale-dependence of the spectrum of $\zeta$. Averaged over
the whole range of scales we need strongly increasing $n$ corresponding
to running $n'\sim 10\mtwo$. Up to this point we assumed nothing about
the origin of $\zeta$.  In Section III we introduce the assumption
 that it  originates from the inflaton
perturbation during slow-roll inflation 
(the standard paradigm). Within this paradigm, the 
 only extant model giving the required running $n'$
is the running mass model, which typically makes $n'$ roughly constant
hence  requiring  $n'\sim 10\mtwo$ on 
cosmological scales.

In Section IV we ask whether such a large value of the running
is still permitted by current
data, thereby updating an earlier work.  We find that it is. 

The question then arises, whether black hole formation can still be
achieved if $n'$ is negligible on cosmological scales, as might be
required by future data.   In Sections V to VIII we show that black hole
formation can indeed be achieved within the standard paradigm.
Along the way, we are led to consider the standard paradigm in more detail
than before. 

  In Section IX we depart from the standard paradigm, by allowing a 
curvaton-type mechanism to contribute to the curvature perturbation.
We show that black hole formation can occur if there is a switch from
the standard paradigm to a curvaton-type paradigm as we go up
in scale, or vice-versa.  We conclude in Section X.

\section{Forming black holes}

\label{sbh}

\subsection{Viable black hole formation}

  The curvature perturbation $\zeta$ is time-independent during any
era when there is a unique relation $P(\rho)$ between pressure and energy
density \cite{lms}. {}From the success of the BBN calculation, we know that
such  is the case to high accuracy a few Hubble times
before  cosmological scales start to come inside the horizon.
On cosmological scales, the  spectrum $\calpz(k)$
is then observed to be about  $(5\times
10^{-5})^2$~\cite{spergel}.\footnote
{The precise number refers to the pivot scale 
defined in Section III. As usual,  $k$ is the coordinate wavenumber so that
$k/a$ is the physical wavenumber, with $a$ the scale factor of the Universe
normalized to 1  at present. The  Hubble parameter is 
$H\equiv \dot a/a$ and horizon entry is defined as $k=aH$.}

  When smaller scales start to come inside the horizon $\calpz(k)$
 could be bigger. To discuss that case, 
recall that the typical value of $\zeta(\bfx)$ in the observable Universe
\cite{mybox},
smoothed on the scale $k\mone$, is of order $\calpz\half(k)$.
 If $\calpz\half(k)$ is bigger than 
 $10\mtwo$ or so, then black holes will form \cite{BBN}
with an abundance that can be ruled
out  \cite{ruleout} by a variety of observations.
A somewhat smaller value, say $\calpz\half\sim 10\mthree$, would give
an abundance whose effect may be observable in the future. We want to
see how such a value may be generated over some range of  $k$.

The spectral index $n$ is defined by
\be
n(k)-1=\frac{d \ln \calpz(k)}{d\ln k}\simeq 
- \frac{d \ln \calpz(N)}{dN}
. \ee
In the  final expression we assume almost-exponential inflation,
with $N(k)$ the number of $e$-folds of inflation remaining after the epoch
of horizon exit $k=aH$ for the scale $k$. We will freely use $N$ as an
alternative variable to $\ln k$.

We take $N_0=50$ unless otherwise stated, where the subscript 0
denotes the epoch of horizon exit for the  present Hubble scale
$k=H_0$. This is the largest observable scale, and smaller scales
leave the horizon 
$\Delta N>0$ $e$-folds later.  For  scales probed by
WMAP and galaxy surveys, $\Delta N\leq 7$~\cite{spergel}.

Until Section IXB  we will assume that the black holes form on the scale
leaving the horizon at the end of inflation, corresponding to $N=0$.
We will take
 the criterion for significant black hole formation to be
$\calpz(0)=10\mthree$. 
Then we need
 $ \ln[\calpz(0)/\calpz(N_0)] \simeq
\ln(10^{-3}/10^{-9}) \simeq 14 $.  With constant $n$ this requires
$n-1\simeq14/N_0\simeq 0.3$. This was compatible with observation for
many years but is now excluded.  

Taking instead $n'\equiv dn/d\ln k$ to be constant  black
hole formation requires  
\be
14=\ln\[\frac{\calpz(0)}{\calpz(N_0)}\] = N_0 (n_0-1) + \frac12 N_0^2 n'
.\ee
Since observation requires $n_0-1\simeq -0.05$ \cite{spergel}, the first
term is  negligible and we need
$n'\simeq 28/N_0^2\simeq 0.01$. As  we will see, this  is compatible with 
 observation.
 At  the end of inflation, 
$n\sub{end}-1\simeq n' \times N_0  \sim  0.5$. This 
 might be regarded as more or less compatible with the
requirement $|n-1|\ll 1$,
that must hold if $\zeta$ at the end of inflation 
is generated by the perturbation in 
a single light field \cite{ss,treview,mycurv}. We are going to assume
such a scenario and therefore rule out $|n\sub{end} - 1|\gsim 1$.

Finally, suppose that $n'$ increases  monotonically as we go down in
scale.  A significant increase would require $|n\sub{end} - 1|\gsim 1$
which we rule out.  A significant decrease would require $n'_0$
significantly bigger than $0.01$, which as we will see would be  in
conflict with observation.

\subsection{The case $\calpz\gg 1$}

We end this section by discussing briefly the case $\calp_\zeta(k)\gg 1$.
If $\zeta$ is generated from the inflaton
perturbation during slow-roll inflation (the standard paradigm),
this is ruled out. The reason is that the regime
$\calpz\gsim 1 $  then corresponds to 
 eternal inflation \cite{eternal},
 whose duration is indefinitely long. Then the slow roll
model has nothing to do with the observed  perturbations, which
instead have to be generated when the eternal inflation is over.

However, such a perturbation  could also
be generated by the perturbation of a curvaton-like field
\cite{curvaton,mycurv}, as one can
readily understand from the non-perturbative formula \cite{lms,deltan,ss}
$\zeta=\delta N$
which makes sense  no matter how big is $\zeta$. In that case, a local
observer would notice nothing amiss before horizon entry, and 
it is not  clear what will happen at horizon entry. 

\section{Running mass model}

Now we assume  that $\zeta$ is generated by the inflaton perturbation
in a single-field slow-roll model. Of the  many such models that have
been proposed,  the only one giving the large   positive running
required for black hole formation is the running mass
model~\cite{running,r2,r3,r4,lgl}\footnote {See for instance
\cite{negrun} and references therein for  models with strong negative
running} . This model invokes softly broken global supersymmetry
during inflation, with a potential \be V=V_0 \pm \frac12 m^2(\phi)
\phi^2 , \ee and a running mass $m^2(\phi)$ whose form is determined
by Renormalization Group Equations (RGE's).

Over the limited range of cosmological scales, $n(N)$  
typically has the  two-parameter form
\be
\frac{n(N) -1}{2} = se^{c(N_0-N)} -c
\label{form} . \ee
With 
\be
c\simeq 10\mone \quad{\rm  to}\quad 10\mtwo
\label{expec} 
\ee
this  gives
\be
n_0= 2(s-c) - 1,  \qquad n_0'= 2sc 
, \ee
whose inverse is
\be
c=-\frac{n_0-1}4 \pm \sqrt{ \(\frac{n_0-1}4 \)^2 +\frac{n_0'}2 }
. \label{inverse} \ee
We see that significant negative running is forbidden;\footnote
{This corrects the relation $n_0'>(n_0-1)^2/4$ given in \cite{r3}.  We
are ignoring the correction to slow roll invoked in part of that work.}
\be
n_0' > - (n_0-1)^2 /8 \sim - 3 \times 10^{-4}
. \ee

Higher derivatives $n^{(m)} \equiv
d^m n/d (-N)^m$ are  suppressed; 
\be
n^{(m+1)} \simeq c n^{(m)}, \qquad m\geq 1
, \ee
but as the form \eqreff{form} is only approximate one should not take
higher derivatives too seriously.

  Going further down in scale, the form of $n(N)$  depends on the assumed 
interactions that determine the RGE's. Typically, $n'$ will increase or
decrease monotonically. As we have seen, black hole formation will then
 need $n'$
to have a roughly constant value, $n' \sim 10\mtwo$, and this
can be achieved  with suitable interactions \cite{lgl}.
{}From \eq{inverse}, $n'\sim 10\mtwo$  corresponds to 
to $c\simeq s\simeq \pm \sqrt{n_0'/2}$, making  
$|c|\sim 10\mone$, in agreement with the expectation
\eqreff{expec}.

\section{Observational bound on the running mass model}

\label{sobs}

 The most recent comparison of the running mass model with 
observation was made in  2004 using  WMAP (year on) and galaxy survey
data available at the time \cite{r4}.
It gave   $n_0'< 0.04$ or so, 
  easily allowing  black hole formation in versions of the model where 
$n'$ does not increase too strongly going down in scale.
In this section we report an update to the earlier bound,
 using year three WMAP data and
more recent galaxy survey results.

In the earlier fit, we took $c$ and $s$ as the parameters to be
fitted,  and only afterward generated  contour plots of $n_0$ versus
$n'_0$.  In the present fit, we instead took  $n_0$ and $n'_0$ as the
parameters  to be fitted. Taking  advantage of the fact that \eq{form}
practically excludes negative $n_0'$, and that it requires $n'$ to
have slow variation, we took $n'$ to be constant and imposed
$n_0'>0$ as a prior. As in the  previous fit we took the  tensor
perturbation to be  negligible since that is a prediction of the
model.  This differs substantially from the method adopted in
\cite{spergel} where the running of the spectral index was let free to
negative values and where tensors were included. In that case a
negative value of the running is obtained, with no running excluded at
the level of $\sim 1 \sigma$ (see
e.g. \cite{spergel,kkmr,easther,finelli,ringe}).  While this approach
is obviously correct when a general set of inflationary models is
considered, it is important to stress  that in our case, where we
don't consider models with $n'<0$, the inclusion of those models could
bias the result towards more restrictive bounds. Moreover, a model
with $n'=0$ gives an acceptable  goodness-of-fit to the WMAP data and
it is therefore  statistically legitimate to assume the prior $n'\ge0$.
 
As is now common practice, we base our analysis on 
Markov Chain Monte Carlo methods making use of the publicly  available
 \texttt{cosmomc} package \cite{Lewis:2002ah}. We sample the following
dimensional set of cosmological parameters, adopting flat priors on them:
the physical baryon and CDM densities, $\omega_b=\Omega_bh^2$ and
$\omega_c=\Omega_ch^2$, the ratio of the sound horizon to the angular diameter
distance at decoupling, $\theta_s$, the scalar spectral index, $n$,
and the optical depth to reionization, $\tau$. We consider purely adiabatic
initial conditions. We choose a pivot scale at $k=0.002 h^{-1}$Mpc.

The MCMC convergence diagnostics are done on $7$ chains applying the
Gelman and Rubin ``variance of chain mean''$/$``mean of chain
variances'' $R$ statistic for each parameter. Our $1-D$ and $2-D$
constraints are obtained after marginalization over the remaining
``nuisance'' parameters, again using the programs included in the
\texttt{cosmomc} package.  Temperature, cross polarization and
polarization CMB fluctuations from the WMAP 3 year
data~\cite{spergel,Page:2006hz,Hinshaw:2006ia,Jarosik:2006ib} are
considered and we include a top-hat age prior  $10 \mathrm{\ Gyr} <
t_0 < 20 \mathrm{\ Gyr}$.

We also consider the small-scale CMB measurements of the CBI 
\cite{2004ApJ...609..498R}, VSA \cite{2004MNRAS.353..732D}, ACBAR
\cite{2002AAS...20114004K} and BOOMERANG-2k2
\cite{2005astro.ph..7503M} experiments. 
We combine the CMB data with the  real-space power spectrum of
galaxies from the 2dF survey \cite{2005MNRAS.362..505C}.
We restrict the analysis to a range of scales over which the
fluctuations are assumed to be in the linear 
regime (technically, $k < 0.2
h^{-1}$~Mpc) and we marginalize over a  bias $b$ considered as an
additional nuisance parameter.

\begin{figure}
    \begin{center}
\includegraphics[width=80mm,clip,keepaspectratio]{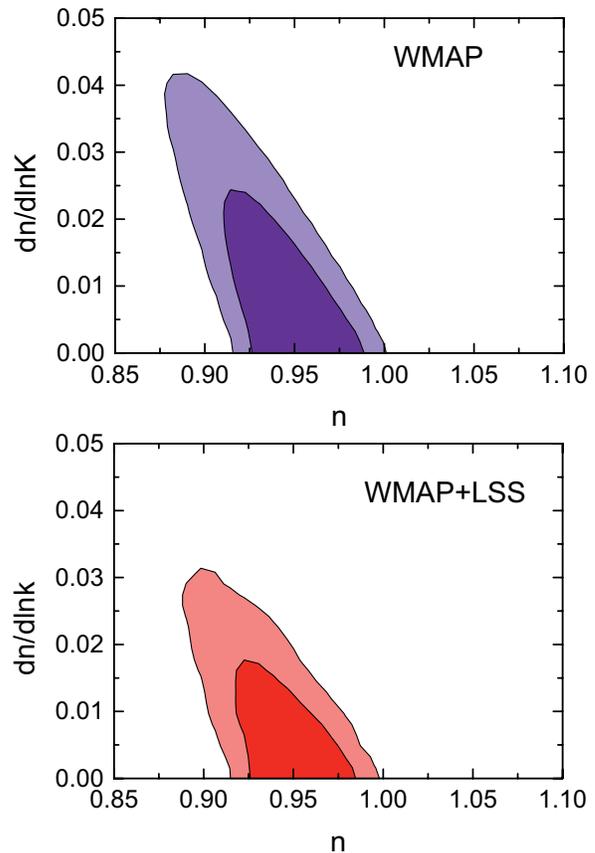}
    \end{center}
    \caption{$68\%$ and $95 \%$ c.l. likelihoods in the $n-n'$ plane from the WMAP data alone
    (Top Panel) and WMAP+LSS (Bottom Panel).}
    \label{fig:bounds}
\end{figure}

In Figure~\ref{fig:bounds} we plot the $68 \%$ and $98 \%$ confidence
levels in the $n$-$n'$ plane for two different choices of our
datasets: WMAP data alone, that should be considered as the most
conservative result, and ``WMAP+LSS'' that includes the remaining CMB
data and galaxy clustering data from 2dF.

As we can see from the Figure, when negative running is not
considered, the data is still in good agreement with a small,
but still non-zero running. When the WMAP dataset is considered
we found a $95 \%$ c.l. upper limit of $n' < 0.039$, while the spectral
index is bound to be $n=0.935_{-0.049}^{+0.039}$ again at 
 $95 \%$. The best fit (maximum likelihood) model 
has a negligible running $n'=0.005$ and $n=0.953$.
When the remaining cosmological data are included, we found a 
stronger bound on running, with
 $n' < 0.026$, and  $n=0.940_{-0.040}^{+0.032}$ at $95 \%$ c.l..
 The best fit (maximum likelihood) model has parameters $n'=0.0026$ 
and $n=0.951$. We conclude that the value $n'\sim 10\mtwo$ required by
the running mass model is still viable.

We have checked that our limits on n' are also  consistent with the
WMAP results  even in the case when a negative running is allowed. The
fitted parametes are in reasonable agreement with the limits we quoted
above, even if slightly more stringent due to the inclusion of
negative running.

\begin{figure}
    \begin{center}
    \includegraphics[width=90mm,clip,keepaspectratio]{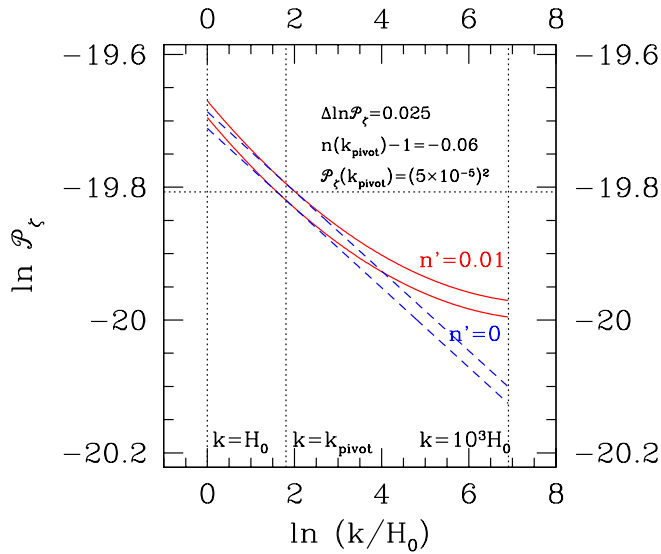}
\end{center}
    \caption{The band  corresponds to the spectrum $\calpz$
versus $\ln k$, with constant slope corresponding to $n=0.94$.
The width of the band corresponds to a fractional uncertainty $0.025$.
The range of $\ln k$ corresponds to the range of cosmological scales,
 explored by observation of the cmb anisotropy and galaxy surveys.
We see that the band is narrow enough to make the 
variation of $\calpz$  significant over the cosmological range.}
    \label{fig:spec}
\end{figure}

In performing this fit, we   chose a pivot point $k=0.002h\Mpc\mone$,
corresponding to $\Delta N=1.8$. Our fitted value $n=0.94$ therefore
corresponds  to $n_0=0.94- 1.8n'$. (Recall that the subscript 0
denotes the scale $k=H_0= 0.00033h\Mpc\mone$.) In Figure
\ref{fig:spec}, we plot  the  shape of the spectrum with (i) no
running and $n=0.95$, (ii),  $n'=0.01$, and  $n=0.95$ at the pivot
point. In the second case, $n-1$  soon climbs to positive values as we
go down below the pivot scale. (The finite width of the band is not
important at this stage, and will be discussed in Section VI.)

\section{Slow roll formalism}

In the future,  observation may require negligible running on
cosmological scales. We have seen that this would not permit black
hole formation if $n'$ increased or decreased monotonically as we go
down in scale, but black hole formation may still be possible with a
more complicated behavior of $n'$.  We are going to  exhibit a couple
of forms of $n(k)$ that would do the  job, and still be compatible
with slow-roll inflation. In order to do that, we need to consider
carefully what the slow-roll approximation involves.~\footnote{We
consider only single-field inflation models. The generation of black
holes has recently been investigated within a double inflation model
with a strong negative running~\cite{Kawaguchi:2007fz}.}

The slow roll formalism is reviewed for 
instance in \cite{treview,book,al,paris}. 
It  starts from the exact 
 Friedmann equation\footnote
{As usual $\mpl=2\times 10^{18}\GeV$  is the reduced Planck mass
 and $V$ is the potential of the inflaton
field $\phi$.}
\be
3\mpl^2H^2 = V(\phi) + \frac12 \dot\phi^2
\label{fried} ,
\ee
and the exact unperturbed field equation
\be
\ddot \phi + 3H\dot\phi + \frac{dV}{d\phi} = 0
\label{field} ,
\ee
from which follow the identity 
\be
\dot H M_{\rm p}^{2}= -\dot\phi^2/2
\label{hident}.  \ee  

In its most basic form, the
  slow-roll approximation  consists  of the two assumptions;
\be
\left| \frac{\dot \phi^2}{H^2 M_{\rm p}^{2}} \right| \ll  1 \qquad
\left| \frac{\ddot \phi}{H\dot\phi} \right| \ll  1
. \ee
These assumptions are usually stated in the equivalent forms
\bea
3\mpl^2H^2 &\simeq & V(\phi) \label{sr1} \\
3H\dot\phi &\simeq & - V'(\phi) 
, \label{sr2} \eea
and they imply $\epsilon\ll 1$ where
\be
\epsilon \equiv \frac12 \mpl^2 \( \frac{V'}{V} \)^2
.\ee

To calculate  the spectrum $\calpz(k)$ one usually considers
 additional
parameters;
\bea
\eta &\equiv&  \mpl^2 \frac{V''}{V} \\
\xi^2& \equiv&  \mpl^4 \frac{V'V'''}{V^2} \\
\sigma^3 &\equiv & \mpl^6 \frac{V'^2 V''''}{V^3} 
. \eea
The last two parameters
can have either sign despite the notation. 

Using \eq{sr2} one finds
\bea
\frac{d\ln H}{dN} &=& \epsilon \label{vflow1} \\
\frac{d(\ln \epsilon)}{dN} &=& - 4\epsilon + 2\eta \label{vflow2} \\
\frac{d\eta}{dN} &=& - 2\epsilon \eta + \xi^2 \\
\frac{d\xi^2}{dN} &=& - (4\epsilon - \eta) \xi^2 + \sigma^3
. \eea
More generally one can define
\be
\alpha_m  \equiv  \mpl^{2m} 
\frac{V'^{m-1} (d^{m+1}V/d\phi^{m+1}) } {V^m} 
,\ee
which satisfy
\be
\frac{d\alpha_m}{dN} = \[ (m-1) \eta - 2 m  \epsilon \] \alpha_m 
+ \alpha_{m+1}
, \label{vflow3} \ee

Assuming that the first derivative of \eq{sr2} is also a valid approximation,
one finds 
\be
\frac{\ddot \phi}{H\dot\phi} = \epsilon-\eta
, \label{secondderiv}\ee
Comparing \eq{sr2} with the exact equation, we see that the fractional
error in \eq{sr2} is ${\cal O} (\epsilon,\eta)$, and so we require
the additional condition  
\be
|\eta|\ll 1
. \label{basichier} 
\ee

Assuming that the curvature perturbation is generated from the vacuum
fluctuation of $\delta\phi$, its spectrum in the slow roll approximation
is given by \cite{spec}
\begin{eqnarray}
    \calpz(k) &=& 
    \frac1{24\pi^2\mpl^4} \frac V \epsilon \[ 1 + {\cal
    O}\(\epsilon,\eta\) \], 
    \label{calpz}
    \\
    &=& 
    \frac{1}{12 \pi^{2}M_{\rm p}^{6}} \frac{V^{3}}
    {\left(3 H \dot{\phi} \right)^{2}} 
    \left[1 + {\cal O}(\epsilon, \eta) \right]
    \label{eq:calpz2}
\end{eqnarray}
The right hand side is evaluated at the epoch of horizon exit.
The displayed uncertainty  takes account of the fractional error in the slow
roll approximation that we just estimated, and first order (linear)
corrections to the calculation of the vacuum fluctuation
  described by the Mukhanov-Sasaki equation \cite{ms,ls,gs,ewanbeyond}
(It does not account
for nonlinear effects, coming from interactions  of  $\delta\phi$. Such 
effects, which would generated non-gaussianity of $\zeta$, are expected to 
be small \cite{maldacena}.

Now  differentiate \eq{calpz} with respect to $\ln k$, using $d\ln k=-dN$
and ignoring the uncertainty. One finds \cite{specind}
\bea
n-1&=& 2\eta-6\epsilon \label{npred}  \\
n' &=& 2\xi^2 + 24\epsilon^2  - 16\epsilon\eta \label{nppred}
. \eea
The  error in $n-1$, coming from the derivative of the 
error in $\calpz$, is ${\cal O}(\xi^2,\epsilon^2,\epsilon\eta)$.
We will 
assume that there is no cancellation between the two terms of \eq{npred},
and make a similar assumption for $n'$ and higher derivatives. Also, we will
assume that $\epsilon$ is negligible compared with $\eta$, $\xi^2$ and 
any  other relevant $\alpha_m$. Then the fractional uncertainty in $n-1$ will
be small if and only if
\be
|\xi^2| \ll  |\eta| \label{h1} 
. \ee
Similarly, the  fractional
error in $n'$ will be small if and only if
\be
|\sigma^3| \ll  |\xi^2| \label{h2}
. \ee
In principle one can go on to calculate higher derivatives of $n$, requiring
a more extended hierarchy 
\be
|\alpha_{m+1}|\ll |\alpha_m|
\label{alphahier}. \ee
{}From \eq{vflow2}, this is equivalent to 
\be
\left| \frac{d^m{\ln \epsilon}}{dN^m}  \right| \ll 
\left| \frac{d^{m-1}{\ln \epsilon}}{dN^{m-1}}  \right|
. \label{ephhier} \ee

We have been exploring the validity of successive derivatives  with
respect to $N$, of the slow-roll approximation \eq{npred} for $n-1$.
Barring cancellations, the validity of these up to a given order will
be equivalent to the validity of  derivatives of the slow-roll
approximation for the field equation \eq{field}, up to one higher
order.  To see this, start with \eq{secondderiv} which expresses the
validity of the  first derivative of \eq{field}.  Put it  into
\eq{field},  and use $\ln(1+x)\simeq x$  to find the approximation 
\be
\ln(|V'|) = \ln(3H|\dot\phi|) +  \frac{\epsilon-\eta}3 . 
\ee 
Assuming that the derivative of the approximation  \eq{secondderiv} is
also valid,  we can differentiate this with respect to $N$: 
\be
\frac{d \ln|V'|}{dN} = \frac{d \ln(3H\dot\phi)}{dN} + \frac{d}{dN}
\(\frac{\epsilon-\eta}3 \). 
\ee
Barring cancellations, the first term on the right hand side is of
order $n-1$ as  is easily seen by comparing it with the derivative of
Eq.(\ref{eq:calpz2}). Therefore, barring cancellations, the validity
of this approximation is indeed equivalent to the validity of the
first derivative of the approximation \eq{npred}, and so on for higher
derivatives.

The equivalence that we saw in the last paragraph means that 
the standard slow-roll approximation for $n-1$, $n'$ and so
on  will be valid, if the second, third and so on 
derivatives of the slow-roll approximation \eq{secondderiv} to the 
exact field equation \eq{field} are valid. 
Reverting to our assumption that $\epsilon$ is negligible, we 
conclude that the hierarchy \eq{h1}, \eq{h2} etc.\ will hold (justifying
the standard formulas for $n-1$, $n'$ etc.) to the 
extent that derivatives of the slow-roll approximation \eq{secondderiv}
hold. 

With the hierarchy in place, one can systematically improve the 
predictions \eqssref{calpz}{npred}{nppred} \cite{ls,gs}. 
The hierarchy is in general satisfied by the running mass model, and
including the 
leading order correction \cite{ls}, the  running mass
prediction \eqreff{form} becomes \cite{r3}
\bea
\frac{n(N) -1}{2} &=&  \( s+ 1.06cs \) e^{c(N_0-N)} -c \\
&\simeq & \(s + 0.50 n' \)  e^{c(N_0-N)} -c  
.\label{improvedn} \eea
Such corrections are usually 
equivalent to a change in parameters whose values are not known
(in this case, a change to $s$),  making them of limited practical
importance.

 Of course, a given inequality in the 
hierarchy will fail for  a few Hubble times if its right hand side
 passes through zero. For instance, \eq{h1} will fail if 
$\eta$ passes through zero. Then, if $\epsilon$ is negligible, $n-1$ will 
pass through zero as well,  and while it is doing so 
the {\em fractional} error in its predicted value will become large.
 According to our fit to the data, $n(N)-1$ will indeed
pass through zero on some scale
 near the bottom
end of the cosmologically accessible range, if $n'$ has a slowly varying
value of order $10\mtwo$.
The passage of $n-1$ through zero need not be a matter of concern, as
 the {\em absolute error} remains the same.
 The running mass 
prediction \eqreff{improvedn} should remain valid even as $n$ passes through
zero.
 
More generally, it could happen  that   derivatives of
\eq{sr2} beyond the first are invalid over an extended range,
 so that the hierarchy fails over an extended range.
 To handle such cases one can use
the exact (at first order) Mukhanov-Sasaki equation or an analytic 
approximation \cite{ewanbeyond}. 

\section{Finite difference version of the spectral index}


\label{ssr}

Although the hierarchy leads to simple and widely-used results, we
have seen that its use may sometimes be problematic and we will see
some more examples of that in the following two sections. For that
reason, we explain in this section how the hierarchy can if necessary
be avoided.

The starting
point is to realise that the  prediction \eq{calpz}, with a suitably
small error, 
 will accurately define the variation of $\calpz$ over a finite range, even 
if the mathematical derivative 
should have large errors (coming for instance from an oscillation
or a break). This is illustrated in Figure \ref{fig:spec}. 

Let us therefore 
redefine  $n-1$ so that it  specifies  a finite difference:
\be
\tilde n-1 \equiv \frac{
\ln \calp_{\zeta 2} -\ln \calp_{\zeta 1}
}
{
\Delta\ln k}
, \ee
where $\Delta\ln k \equiv \ln k_2 - \ln k_1$ and $\calp_{\zeta i}
\equiv \calpz(k_i)$.
To the extent that  observational bounds on the variation of $n(k)$ are quite
weak, this finite difference is really about all that observation can determine
at present, with $\Delta\ln k\simeq 7$ or so.

The error in  $\tilde n-1$ generated by a fractional 
 error $x$ in the prediction \eqreff{calpz}  will be at most of order
\be
\delta (\tilde n-1) \simeq  \frac{x}{ \Delta\ln k}
. \ee
Let us assume  $x\simeq 0.025$, corresponding to \eq{calpz}
with  $\epsilon\ll |\eta|\simeq 0.05$ (the observed value of
$|n-1|$).
Then the error in the theoretical
prediction will be small, if the prediction satisfies
\be
|\tilde n-1| \gg \frac{x}{\Delta\ln k} \simeq 0.004
. \ee
As illustrated in Figure \ref{fig:spec}, this is very well satisfied
if $\tilde n-1$ has the observed value $\simeq -0.05$. 

We can go a bit further, to consider
 a finite-difference version of the running;
\bea
\tilde n'& \equiv & \(
\frac{\ln \calp_{\zeta 1} -\ln \calp_{\zeta 3}}{\Delta\ln k/2}
-\frac{\ln \calp_{\zeta 3} -\ln \calp_{\zeta 2}}{\Delta\ln k/2}  
\)/\(\Delta \ln k/2 \) \\
&=&  \( \frac{\ln \calp_{\zeta 2} -2\ln \calp_{\zeta 3} +\ln \calp_{\zeta 1}}
{(\Delta \ln k)^2/4} \)
, \eea
with $2\ln k_3 \equiv \ln k_1 + \ln k_2$. The error in $\tilde n'$ generated
by the error $x$ in the prediction will be at most of   order\footnote
{The factor 10 accounts roughly for the $1/4$ in the denominator and
the four terms of the numerator.}
\be
\delta \tilde n' \sim \frac{10x}{(\Delta \ln k)^2}  
. \ee
The error  will be small if the prediction satisfies
\be
|\tilde n'| \gg \frac{10x}{(\Delta\ln k)^2} \simeq 5\times  10\mthree
, \ee
where we again set $x=0.025$ as an illustration.
Taking account of the uncertainties, the prediction for the
finite-difference version of the 
running may be valid if $n'\sim 10\mtwo$.

\section{Flow equations}

In the above analysis we worked directly with the potential. A different
approach works initially with the field $\phi(t)$, connecting only later
with the potential. The starting point is \eq{hident}, providing
a parameter $\eph$  which may be
written in various forms;
\begin{eqnarray}
    \eph &\equiv& -\frac{\dot H}{H^2}
    = \frac{d(\ln H)}{dN} \nonumber \\
    &=& \frac1{2\mpl^2}  \( \frac{d \phi}{dN} \)^2
    =2\mpl^2  \( \frac1H \frac{dH}{d\phi} \)^2, 
    \label{neweqcond}.
\end{eqnarray}
Its derivatives satisfy the exact set of equations
\bea
\frac{d(\ln \eph)}{dN} &=& - 2(\eph   + \delta_1) \\
\frac{d\delta_m}{dN} &=& ( \delta_1 - m \eph )\delta_m + \delta_{m+1}
, \label{ewanflow1}\eea
where
 \cite{gs,ewanbeyond} 
\be
\delta_m\equiv d^m\phi/dt^m/H^m\dot\phi
. \label{ewanflow2}\ee

Equivalently, one can use $\phi$ instead of $N$ as the variable. Then
\cite{flow1}
\bea
\frac{d(\ln (\eph)}{d\phi} &=& -2(\eph - \beta_1) \\
\frac{d\beta_m}{d\phi} &=& \[ (m-1) \beta_1 - m\eph \] \beta_m + \beta_{m+1}
, \label{floweq} \eea
where \cite{flow2}
\be
\beta_m \equiv  \(
\frac{2\mpl} {H } \)^m
\( \frac{dH}{d\phi} \)^{m-1}
\frac{d^{m+1} H}{d\phi^{m +1}}
\label{flowedef}, \ee
These are referred to as flow equations.

The flow equations (equivalently \eqs{ewanflow1}{ewanflow2}) resemble
\eqss{vflow1}{vflow2}{vflow3} but are exact. 
Slow roll with the potential hierarchy \eqreff{alphahier},
up to $m=M$, is obtained if there is a hierarchy\footnote
{The  stronger hierarchy 
$|\delta_{m+1}|^{1/(m+1)}\lsim |\delta_{m}|^{1/(m)}$ is sometimes considered
(equivalently $|\beta_{m+1}|^{1/(m+1)}\lsim |\beta_{m}|^{1/(m)}$. It implies
the potential hierarchy $|\alpha_{m+1}|^{1/(m+1)}\lsim |\alpha_{m}|^{1/(m)}$,
which is satisfied by a wide class of potentials but not by the running
mass potential.}
\be
|\delta_{m+1}| \ll |\delta_m|
, \label{deltahier}\ee
or equivalently
\be
|\beta_{m+1}| \ll |\beta_m|
\label{betahier}
\ee
up to $m=M+1$. Following \cite{flow2} 
one might call this the `Hubble hierarchy', as opposed to the
`potential hierarchy' \eqreff{alphahier}.

Conversely, if  the potential  hierarchy is satisfied up to $m=M$, then
 one can expect the solution $\phi(t)$ to satisfy the hierarchy
\eqreff{betahier} (equivalently \eqreff{deltahier}) up to $m=M+1$, 
 at least for low $M$. 
This is because the slow-roll approximation \eqreff{sr2} is known
to be a strong attractor for a wide range of initial conditions. 
As with the potential hierarchy, the Hubble hierarchy will fail briefly 
if a parameter ($\delta_m$ or $\beta_m$ passes through zero, and it might not
be valid at all.

\section{Two forms for the potential}

Now we consider forms of the potential, which would permit slow
roll inflation leading  to black
hole formation, and be consistent with a negligible value of $n_0'$.
A common procedure for generating potentials consistent with assigned
values of (say) $n_0$ and $n_0'$ uses the 
flow equations. The equations are numerically integrated with 
an initial hierarchy imposed such as $|\beta_{m+1}|/|\beta_m| < 1/5$
\cite{ce} or $1/10$ \cite{flow1}. This procedure is quite complicated, and
will obviously miss
potentials violating the initial hierarchy as discussed at the end of
the previous section.\footnote
{The flow equations were used in \cite{ce} to search for potentials
consistent with black hole formation but none were found. The authors
concluded that ``... it seems extremely unlikely to us that primordial
black holes formed as a result of inflationary dynamics''. It was the 
discrepancy between this result and the earlier positive conclusion
of \cite{lgl} that motivated the present investigation. We suppose that it
is caused by the use in \cite{ce} of the flow equations and the  hierarchy.}

Our procedure will be to simply  specify  suitable 
forms for  $d(\ln\epsilon)/dN)$. 
(It  resembles the one advocated in \cite{andrewflow}.)
{}From these the 
potential can be constructed using 
\begin{eqnarray}
\epsilon(N) &=& \epsilon(N_0) \exp \[-\int^{N_0}_N  \frac{d\ln \epsilon}{dN}
 \] \\
H(N) &=& H(N_0)  \exp \[ -\int^{N_0}_N \epsilon(N)  dN \] \\
\phi(N_0)-\phi(N) &=& \mpl \int^{N_0}_N \sqrt{2\epsilon(N)} dN  
 \label{first} \\
V(\phi) &=&  3\mpl^2H^2(\phi) \label{last}.
\end{eqnarray}
The value $H(N_0)$ is determined once the inflation scale 
$V(N_0)$ is set, and then
 $\epsilon(N_0)$  is  obtained from \eq{calpz}
using   the observed value  $\calpz(N_0)  =
(5\times 10^{-5})^2$.

  To keep things simple
we  focus on small-field inflation, which  corresponds to  $\epsilon$
far  below 1. To 
be precise, we assume $\epsilon \ll 1/N_0$ for $0<N<N_0$,
corresponding  to tensor fraction $r\ll 16/N_0 \simeq 0.03$.  Then $V$
can be taken to be constant, and black hole formation requires \be
\Delta\ln\epsilon \equiv \ln[\epsilon(N_0)/\epsilon(0)] \simeq
\ln[\calpz(0)/\calpz(N_0)] \simeq 14  . \label{bh14} \ee

The  predictions for the spectral index and its running are then 
\bea
\frac{n-1}2&=&\eta \equiv 
\mpl^2 \frac{V''}{V} ~\left(\simeq \frac12 \frac{d\ln \epsilon}{dN} \right)\\
\frac{n'}2 &=&  \xi^2 \equiv  \mpl^4 \frac{V' V'''}{V^2} 
        ~\left(\simeq \frac12 \frac{ d^2\ln \epsilon}{dN^2} \right) 
\eea

\begin{figure}
    \begin{center}

    \vspace{0.5cm}

    \includegraphics[width=80mm,clip,keepaspectratio]{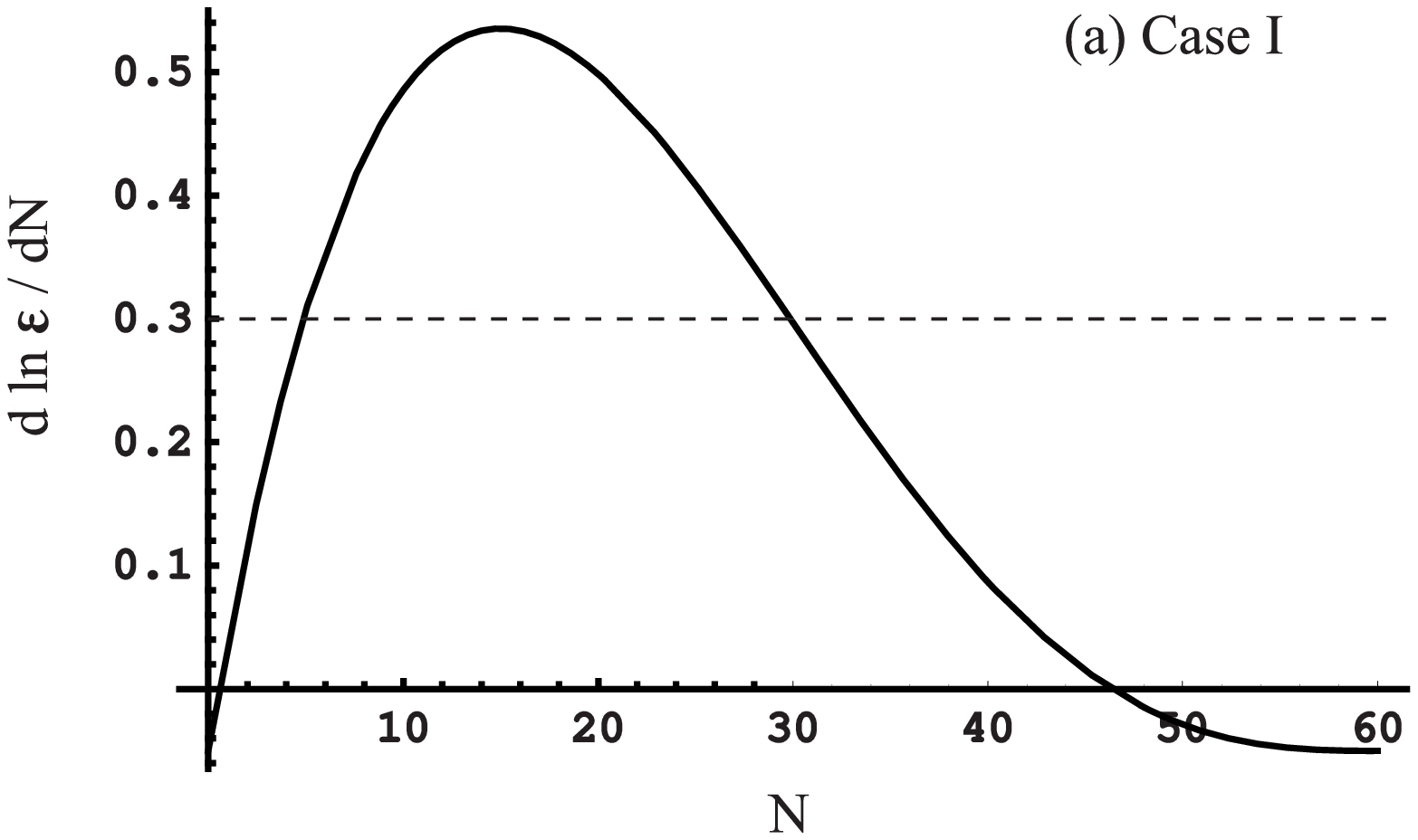}
    \vspace{0.5cm}

    \includegraphics[width=80mm,clip,keepaspectratio]{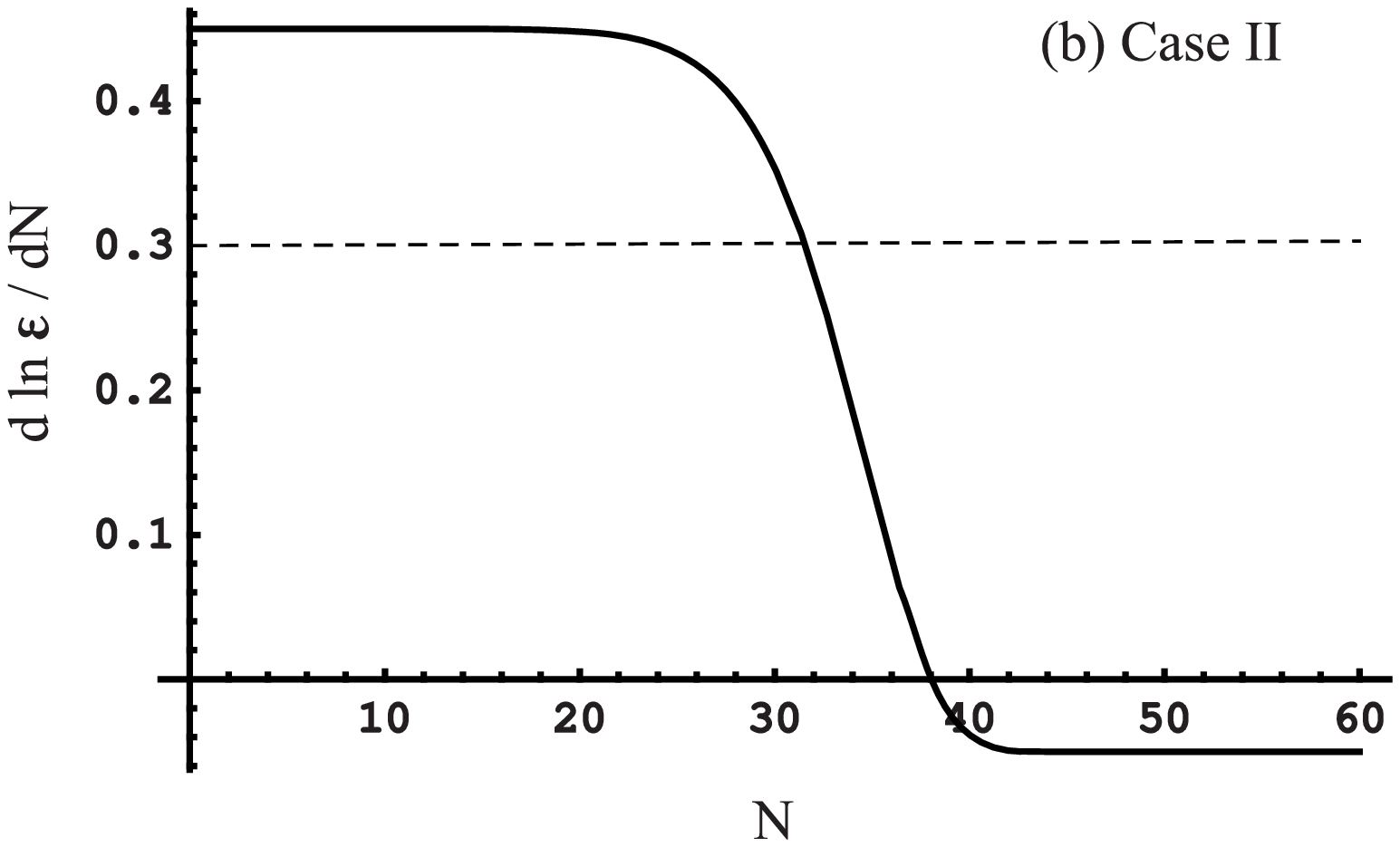}
    \end{center}
    \caption{Schematic pictures of functional form of y(N) in Case I  (top)
    and  Case II (bottom), respectively. For reference we also plot the
    constant case, $n = 0.3$ (see the text).  }
    \label{fig:polynomial}
\end{figure}

\begin{figure}
    \begin{center}
    \vspace{0.5cm}

\includegraphics[width=80mm,clip,keepaspectratio]{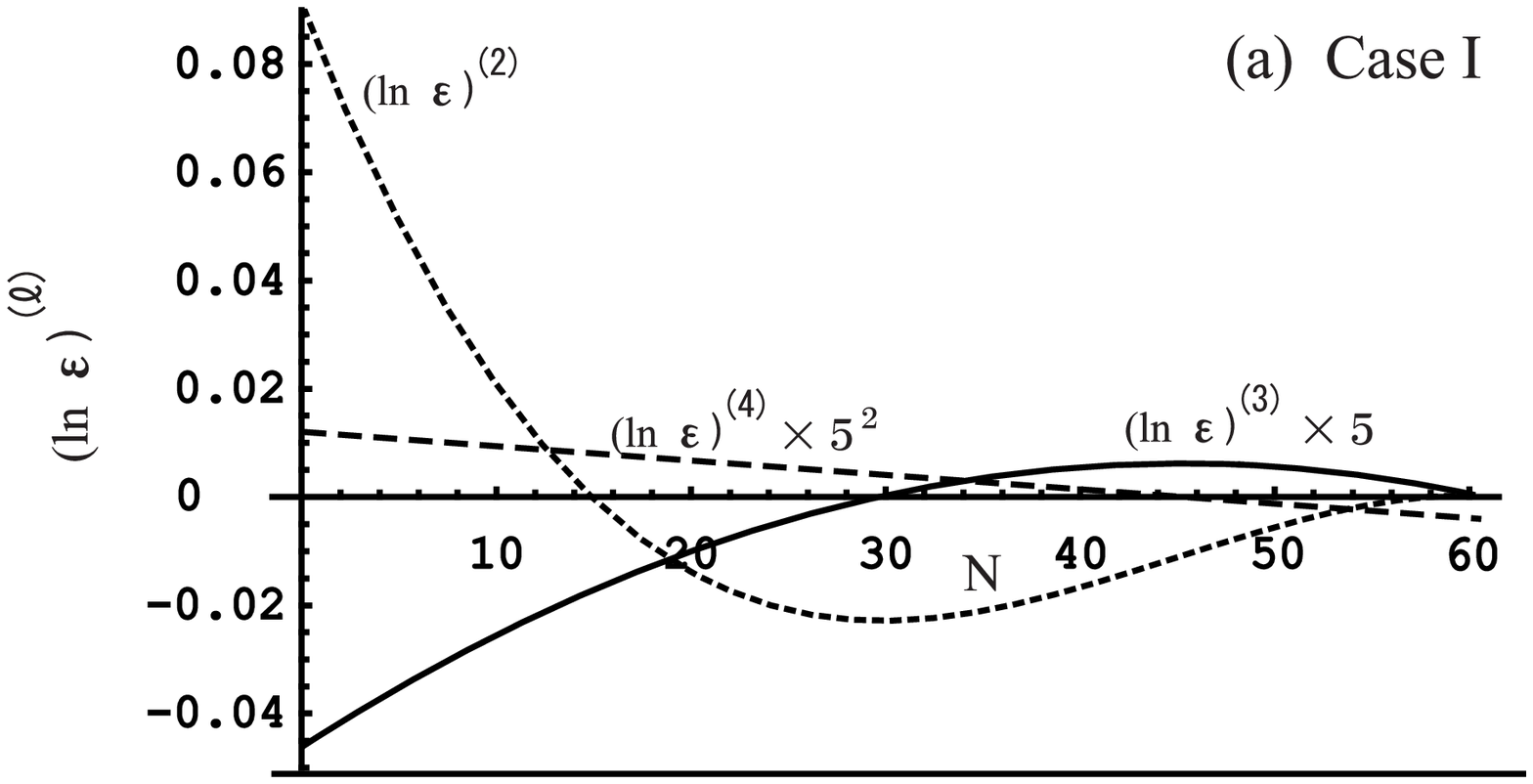}
    \vspace{0.5cm}

\includegraphics[width=80mm,clip,keepaspectratio]{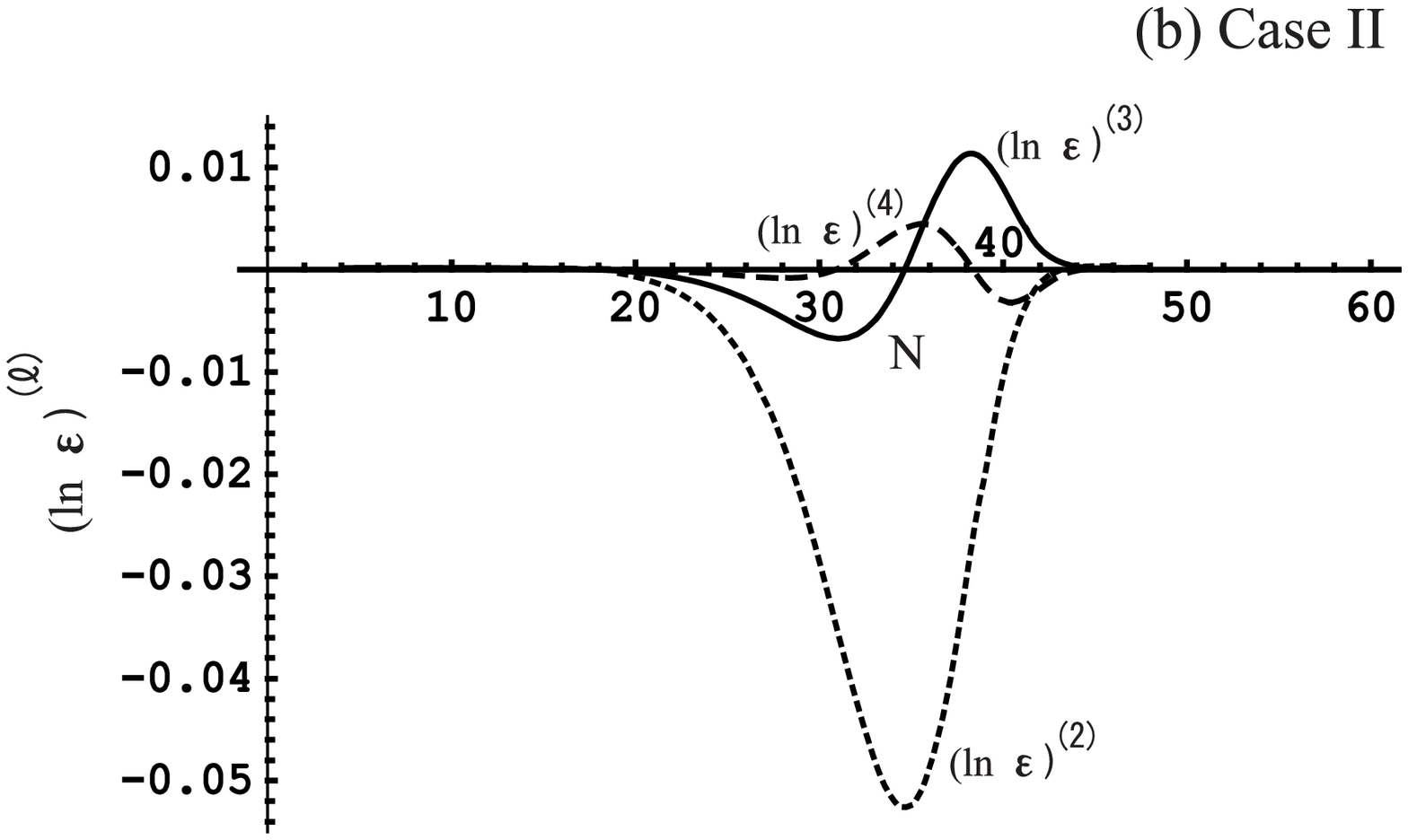}
    \end{center}
    \caption{Derivatives of $d (\ln \epsilon)/dN$ with respect to $N$
    for Case I  (top)  and  Case II (bottom), respectively. Here we
    plot the higher derivatives (ln $\epsilon)^{(\ell)} =  d^{(\ell)}
    (\ln \epsilon)/dN^{(\ell)}$ for $\ell$ = 2, 3, and 4.}
    \label{fig:derivs}
\end{figure}

We have considered    following two  forms.
\begin{eqnarray}
    \label{eq:polynomial}
    \frac{d(\ln \epsilon)}{dN} = B \left( \frac{N}{N_0}\right)^{p} 
     \left( 1 - \frac{N}{N_0} \right)^{q}
      - D  : {\rm Case~I}, \nonumber \\
\end{eqnarray}
and
\begin{eqnarray}
    \label{eq:exponential}
\frac{d(\ln \epsilon)}{dN} = B \exp \left[ - \left( \frac{N}{N_0- \Delta
    N - A}\right)^{q} 
    \right] - D  :  {\rm Case~II}. \nonumber \\
\end{eqnarray}
In Fig.~\ref{fig:polynomial} we plot the schematic pictures in
 case I  (top)
and  case II (bottom), respectively.
In figure \ref{fig:derivs} we show the derivatives
of $d\ln \epsilon/dN$ with respect to $N$. The hierarchy \eqreff{ephhier}
 is in 
general respected except where the denominators pass through zero.

We impose the observational constraints on $\calpz$, $n$ and $n'$. To
be on the safe side we also impose  a rough finite-difference version
of the constraint on $n'$ in the following way. The  WMAP data spans a
range roughly $\ell \sim {\cal O}(1)$ -- $ {\cal O}(10^{3})$,
corresponding to $\Delta N \sim 7$ $(\equiv \Delta N$, and
$n$ derived from that data has an error of about $0.1$. Therefore, we
require that $n$ should change by less than  0.02 in the range $N_0$
to $N_0- \Delta N$.  We have checked that the condition
\eqreff{bh14} needed for the PBH production  is satisfied with all of
the observational constrains
for $p = 1$, $q=3$, $B \simeq 5.5$, $D \simeq
0.05$, $N_0 = 60$ and $\Delta N = 5$ for Case I,
and $q=10$, $A=5$, $B \simeq 0.5$, $D \simeq 0.05$, $N_0 =
50$ and  $\Delta N = 10$  for Case II. For the detailed
numerical values of $n$ and $n'$, see Table~\ref{table:nnprime}.
It is clear that with a parameterisation like the one in Figure 
\ref{fig:polynomial}, we can make $n$ practically constant over the
range $\Delta N\sim 10$ of scales probed by large-scale observations.

\begin{table}
    \begin{center}
        \begin{tabular}{lccccc}
            \hline \hline
            & $n(N_0)$ & $n(N_0-\Delta N$) & 
            $n'(N_0)$ & 
            $n'(N_0-\Delta N)$ & \\
            \hline
            Case I & 0.9500 & 0.9529 & 0 & 0.0017 & \\
            Case II & 0.9500 & 0.9511 & $1.4942 \times 10^{-15}$ &
             $5.9700  \times 10^{-16}$ & \\
            \hline \hline
        \end{tabular}
        \caption{$n$ and $n'$ at $N=N_0$ and $N=N_0 -\Delta N$ in  
Case I and Case II.}
        \label{table:nnprime}
    \end{center}
\vspace{-1cm}
\end{table}

\begin{figure}
    \begin{center}
\includegraphics[width=80mm,clip,keepaspectratio]{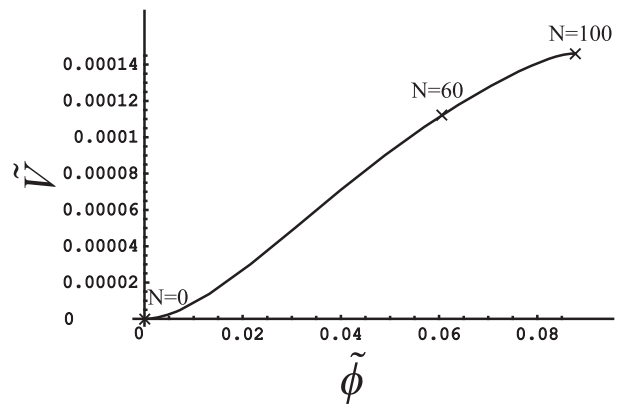}
    \end{center}
    \caption{Form of the potential $V$ as a function of the field $\phi$.
    The horizontal axis is the normalized value of $\phi$,
    $\tilde{\phi} = \left[\frac{ \phi - \phi_{0} }{M_p}\right]/
    \left[\frac{V_{0}^{1/4}}{M_p}/10^{-3} \right]^2$ with $\phi_{0}
    \equiv \phi(0)$ and $V_{0} \equiv V(\phi_{0})$. The vertical axis
    means $\tilde{V} = \left[ \frac{V}{V_{0}}-1 \right] /
    \left[ \frac{V_{0}^{1/4} }{M_{p}}/ 10^{-3} \right]^{4}$.}
    \label{fig:form}
\end{figure}

For the parameterisation I, the potential has a scaling $V_0\propto
(\phi-\phi_0)^2$. We plot it in Figure \ref{fig:form}.  Note that
$V_{0}^{1/4}/M_{p}$ should be less than $\sim 10^{-3}$ for the
slow-roll inflation $\epsilon \ll 1/N_0$ and $\phi < M_{p}$. This
shape is similar to those in some classes of  hilltop inflation
models~\cite{hilltop}.

\begin{figure}
    \begin{center}
    \vspace{0.5cm}
    \includegraphics[width=80mm,clip,keepaspectratio]{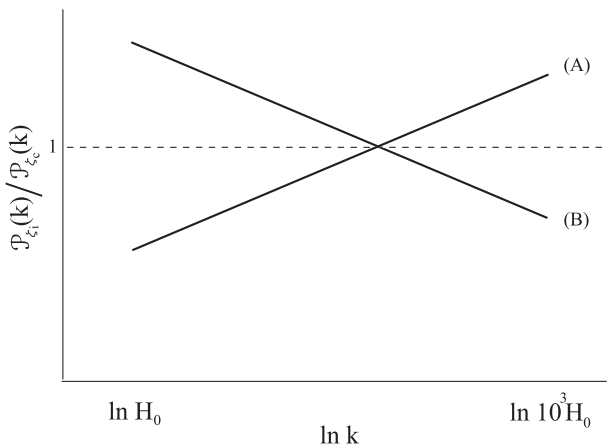}
    \end{center}
    \caption{Two scenarios for generating black holes using a curvaton-type
paradigm.}
    \label{fig:pipc}
\end{figure}


\section{Black hole formation in a curvaton-type paradigm}

\label{scurv}
The inflaton contribution  $\zeta_i(k)$ is time-independent, and is the only
one present at horizon exit. Subsequently though, the contribution 
$\zeta_c(t,k)$
of some
other (curvaton-type) field could grow and become dominant.
(See \cite{mycurv} for a discussion of the possibilities with extensive
references.)
Eventually $\zeta(t)$ will level out to the observed value;
\be
\zeta(k) = \zeta_i(k) + \zeta_c(k)
, \ee
where the last term is the eventual time-independent value of the 
curvaton-type contribution. In an obvious notation, the observed
spectrum is now
\be
\calpz = \calp_i + \calp_c
, \ee
and the spectral index is
\be
n-1 = f_i (n_i-1) + f_c (n_c-1)
, \ee
where $f_i=\calp_i/\calpz$ and $f_c=\calp_c/\calpz$.
We will consider two different possibilities for the ratio
$\calp_i(k)/\calp\sub c(k)$, illustrated in Figure \ref{fig:pipc}.

\subsection{Black holes from the inflaton perturbation}

We first assume that $f_c\ll 1$ at the end of inflation so that black
holes are generated by the inflaton perturbation, but that $f_i\ll 1$
while cosmological scales leave the horizon.
To agree with observation, we will demand at $N=N_0$
\begin{eqnarray}
    \label{eq:nceqni}
    n_{c} -1  \simeq n -1 \simeq 0.05,
\end{eqnarray}
and
\be
f_i |n_i-1| < 1\times 10\mtwo.
 \label{req1} 
\ee
To form black holes, \eq{bh14} becomes
\bea
14 &=& \ln[ \calpz(0)/\calp_c(N_0) ] \\
&=&  \ln[ \calpz(0)/\calpz(N_0) ] + \ln[f_i(0) ] \\
&=& (n_i-1) N + \ln(f_i(0)), 
\label{req2} 
\eea
where we set $n_i'=0$ to get the last line.  These requirements are
satisfied with, for example, 
$f_i(0)\simeq 10\mone$ and $n_i\simeq 1.4$, and the
observational bound on the running imposes no further constraint.

The required spectral index $n_i$ corresponds to $\eta=0.2$ which is
more or less compatible with the slow-roll requirement $|\eta|\ll 1$.
In the context of supergravity such a value is more natural than the
small value $\eta=-0.025$ required to fit observation. This looks
promising for black hole formation, but we have to remember that in a
curvaton type model the prediction for $n$ becomes \cite{mycurv} \bea
n-1 &=&  2\eta_{\sigma\sigma} - 2\epsilon \\
\eta_{\sigma\sigma}&\equiv& -\frac1{3H_*^2}\frac{\pa^2 V}{\pa
\sigma^2}  . \eea If $\epsilon$ is negligible as we are assuming, this
requires $\eta_{\sigma\sigma}=-0.025$ which may difficult to achieve
since $\sigma$ will tend to roll away from any maximum of its
potential \cite{mycurv}. In a curvaton type model it  may therefore be
more attractive \cite{al} to take $\eta_{\sigma\sigma}$ negligible and
$\epsilon=0.025$, but we have not explored that option.

\subsection{Black holes from the curvaton-type contribution}

Now we suppose that the roles of the inflaton and the curvaton are reversed,
so that the inflaton generates the observation curvature perturbation 
while the curvaton perturbation generates black holes. In this case
black hole generation occurs only when the curvaton-type mechanism
operates which will usually be long  after inflation is over.

In this  scenario, we have to modify \eqst{req1}{req2} 
by interchanging $i$ and $c$, {\em and} replacing the epoch $N=0$
by an earlier epoch $N\sub{curv}$. This is the epoch at which the
scale leaves the horizon, that is entering the horizon when the 
curvaton mechanism operates.  To achieve black hole formation we will
therefore generally need  $n_c> 1.4$, but that need not be a problem.
Indeed, within the context of supergravity a value $n_c$ significantly bigger
than 1 is expected \cite{mycurv}, just as it is for $n_i$. This, our third
scenario for generating black holes, therefore seems at least as good
as the other two.

A particularly interesting possibility in this case, is that the
curvature perturbation generating the black holes could easily be
highly non--gaussian, to be precise the square of a gaussian 
quantity \cite{mycurv}. This would not make much of a change \cite{hidalgo}
in the
magnitude of $\calpz$ needed to generate black holes (upon which our
estimates are based) but it could alter the predicted shape of the 
black hole mass function. 

\section{Conclusion}

\label{sconc}

The possibility of primordial black hole formation at the end of inflation
has a long history, which was long overdue for an update. 
The update is needed partly because observation now requires 
{\em on cosmological scales} a tilt far below
$0.3$ (and with negative sign) and not too much running. 
It is also needed because the original
paradigm, that the inflaton perturbation is entirely responsible for the
curvature perturbation, is now only one possibility.

According to the standard paradigm, the curvature perturbation is generated
during slow roll inflation 
from the vacuum fluctuation of the inflaton field. 
Within this  paradigm, the running-mass model provides a
well-motivated way of achieving black hole formation. To form black holes,
the  model  probably requires strong running on cosmological scales,
$n'\sim 0.01$.

Such  running is allowed by the data. If a value $n' \sim 10\mtwo$
on cosmological scales  is ruled out in the future,  $n'$ will still
have to increase to $\gsim 10\mtwo$ in order to  form black holes. We
saw that this may be achieved within the standard paradigm by a
suitable potential. Alternatively, it might be achieved by a switch
from the standard paradigm  to a curvaton-type paradigm,  or by a
switch from the curvaton-type paradigm to the standard paradigm.
 
{}For observers, we would like to re-iterate and earlier conclusion
\cite{paris}, that 
an upper bound $n'<10\mthree$, or detection, would have important implications
and is a very worthwhile goal.

\section{Acknowledgements}
We thank Will Kinney and Andrew Liddle for comments on the draft of this
paper. The research at Lancaster is supported by PPARC grant
PP/D000394/1   
and by EU grants MRTN-CT-2004-503369 and
MRTN-CT-2006-035863.



\end{document}